\documentclass[conference,a4paper]{IEEEtran}
\IEEEoverridecommandlockouts
\usepackage[utf8]{inputenc}

\usepackage[mathscr]{euscript} 
\usepackage[nolist]{acronym} 
\usepackage[normalem]{ulem} 
\usepackage[nospace,noadjust]{cite} 
\usepackage{algpseudocode}
\usepackage{amsmath,amssymb,amsfonts} 
\usepackage{amsthm} 
\usepackage{array}
\usepackage{float}
\usepackage{flushend} 
\usepackage{gensymb}
\usepackage{mathrsfs} 
\usepackage{multirow} 
\usepackage{soul} 
\usepackage{textcomp} 
\usepackage{xcolor}
\usepackage{color}
\usepackage{graphicx}
\usepackage{enumitem}
\usepackage[inkscapeformat=png]{svg}

\usepackage{tikz}
\tikzset{
  font={\fontsize{9pt}{12}\selectfont}}
\usetikzlibrary{plotmarks}
\usetikzlibrary{intersections,backgrounds, patterns}
\usetikzlibrary{patterns,shapes.arrows}

\usepackage{pgfplots}
\pgfplotsset{compat=newest} 
\usepgflibrary{arrows}
\usepgfplotslibrary{groupplots,dateplot}
\pgfplotsset{every axis/.append style={
	scaled x ticks = false,
	label style={font=\footnotesize},
	tick label style={font=\footnotesize},
	tick scale binop=\times}
}
\pgfkeys{/pgf/number format/.cd,
1000 sep={},
}

\usepackage[cmintegrals]{newtxmath} 
\usepackage[font=footnotesize]{subcaption}
\usepackage[font=footnotesize]{caption}
\usepackage[nolist]{acronym}
\usepackage{enumitem}
\DeclareUnicodeCharacter{2212}{−}

\definecolor{mycolor1}{RGB}{19, 133, 189}
\definecolor{mycolor2}{RGB}{230, 112, 32}
\definecolor{mycolor3}{RGB}{130, 173, 98}
\definecolor{mycolor4}{rgb}{0.49412,0.18431,0.55686}%
\definecolor{myhist}{RGB}{73, 80, 87}
\definecolor{hgreen}{rgb}{0, 0.5, 0}

\def\BibTeX{{\rm B\kern-.05em{\sc i\kern-.025em b}\kern-.08em
    T\kern-.1667em\lower.7ex\hbox{E}\kern-.125emX}}

\begin{document}

\title{Distributed Deep Learning for Modulation Classification in 6G Cell-Free Wireless Networks
\thanks{The work of Hazem Sallouha was funded by the Research Foundation – Flanders (FWO), Postdoctoral Fellowship No. 12ZE222N.\\This work is supported by the 6G-Bricks project under the EU’s Horizon Europe Research and Innovation Program with Grant Agreement No. 101096954. }}


\author{\IEEEauthorblockN{Dieter Verbruggen, Hazem Sallouha, and Sofie Pollin}
\IEEEauthorblockA{\textit{Department of Electrical Engineering (ESAT) - WaveCoRE}\\
KU Leuven, 3000 Leuven, Belgium\\
E-mail:\{dieter.verbruggen, hazem.sallouha, sofie.pollin\}@kuleuven.be}}
\maketitle
\begin{acronym}[HBCI]
\acro{amc}[AMC]{Automatic Modulation Classification}
\acro{6g}[6G]{6th Generation}
\acro{nr}[NR]{New Radio}
\acro{oran}[ORAN]{Open Radio Access Network}
\acro{rf}[RF]{Radio Frequency}
\acro{ru}[RU]{Radio Unit}
\acro{du}[DU]{Distributed Unit}
\acro{cu}[CU]{Central Unit}
\acro{iq}[IQ]{todododod}
\acro{cnn}[CNN]{Convolutional Neural Network}
\acro{ap}[AP]{access point}
\acro{ue}[UE]{user equipment units}
\acro{iq}[IQ]{In-phase/Quadrature}
\acro{fft}[FFT]{Fast Fourier Transform}
\acro{cp}[CP]{Cyclic Prefix}
\acro{bpsk}[BPSK]{binary phase shift keying}
\acro{qpsk}[QPSK]{quadrature phase shift keying}
\acro{qam}[QAM]{quadrature amplitude modulation}
\acro{egc}[EGC]{Equal Gain Combining}
\acro{snr}[SNR]{signal-to-noise ratio}
\acro{lb}[LB]{Likelihood-Based}
\acro{fb}[FB]{Feature-Based}
\acro{dl}[DL]{Deep Learning}
\acro{dnn}[DNN]{Deep Neural Networks}
\acro{oran}[O-RAN]{Open Radio Access Network}
\acro{flop}[FLOP]{floating-point operation}
\acro{mflop}[MFLOP]{Million FLOP}
\end{acronym}
\begin{abstract}
In the evolution of 6th Generation (6G) technology, the emergence of cell-free networking presents a paradigm shift, revolutionizing user experiences within densely deployed networks where distributed access points collaborate. However, the integration of intelligent mechanisms is crucial for optimizing the efficiency, scalability, and adaptability of these 6G cell-free networks. One application aiming to optimize spectrum usage is Automatic Modulation Classification (AMC), a vital component for classifying and dynamically adjusting modulation schemes. 
This paper explores different distributed solutions for AMC in cell-free networks, addressing the training, computational complexity, and accuracy of two practical approaches. The first approach addresses scenarios where signal sharing is not feasible due to privacy concerns or fronthaul limitations. Our findings reveal that maintaining comparable accuracy is remarkably achievable, yet it comes with increased computational demand. 
The second approach considers a central model and multiple distributed models collaboratively classifying the modulation. This hybrid model leverages diversity gain through signal combining and requires synchronization and signal sharing. The hybrid model demonstrates superior performance, achieving a 2.5\% improvement in accuracy with equivalent total computational load. Notably, the hybrid model distributes the computational load across multiple devices, resulting in a lower individual computational load.  



\end{abstract}


\begin{acronym}[HBCI]
\acro{amc}[AMC]{Automatic Modulation Classification}
\acro{6g}[6G]{6th Generation}
\acro{nr}[NR]{New Radio}
\acro{oran}[ORAN]{Open Radio Access Network}
\acro{rf}[RF]{Radio Frequency}
\acro{ru}[RU]{Radio Unit}
\acro{du}[DU]{Distributed Unit}
\acro{cu}[CU]{Central Unit}
\acro{iq}[IQ]{todododod}
\acro{cnn}[CNN]{Convolutional Neural Network}
\acro{ap}[AP]{access point}
\acro{ue}[UE]{user equipment units}
\acro{iq}[IQ]{In-phase/Quadrature}
\acro{fft}[FFT]{Fast Fourier Transform}
\acro{cp}[CP]{Cyclic Prefix}
\acro{bpsk}[BPSK]{binary phase shift keying}
\acro{qpsk}[QPSK]{quadrature phase shift keying}
\acro{qam}[QAM]{quadrature amplitude modulation}
\acro{egc}[EGC]{Equal Gain Combining}
\acro{snr}[SNR]{signal-to-noise ratio}
\acro{lb}[LB]{Likelihood-Based}
\acro{fb}[FB]{Feature-Based}
\acro{dl}[DL]{Deep Learning}
\acro{dnn}[DNN]{Deep Neural Networks}
\acro{oran}[O-RAN]{Open Radio Access Network}
\acro{flop}[FLOP]{floating-point operation}
\acro{mflop}[MFLOP]{Million FLOP}
\end{acronym}
\section{Introduction}
\subsection{Motivation}

The technological progress of future \ac{6g} mobile networks creates a groundbreaking paradigm shift in the wireless communications landscape towards more densely distributed deployments and embedded use of Artificial Intelligence (AI) \cite{6gai}. These future networks will use multiple frequency bands together with a wide range of different numerologies and modulations at the physical layer, taking advantage of a wide range of possible distributed precoding, beamforming, or receive combining methods \cite{techniques ,distcellfree}. Advanced sensing and low-overhead control solutions will be needed to optimally allocate resources in such distributed networks.


In the context of 6G, we now see a trend towards cell-free distributed deployments \cite{distributeddeploy}. To overcome path-loss and give homogenous \ac{snr} for all user locations, the dense distribution of transmitters and receivers is a good strategy, giving what is known as \emph{favorable path-loss conditions} \cite{pathloss}. To this end, antenna-to-user allocation strategies are needed to optimally combine signals from multiple locations. Hence, distributing the computation workload between different network components becomes essential to ensure the scalability of distributed networks. 

A popular practical variant of cell-free networks is based on the \ac{oran} architecture \cite{cell-free-oran}, in which the distribution of the computational workload is a crucial aspect. This distribution is manifested in the disaggregation of the network architecture into a \ac{cu}, \ac{du} and \ac{ru}. In this paper, we adopt this terminology, assigning the RU responsibility for lower-level physical layer processing, while the DU handles the final processing of the physical layer. The CU is outside the scope of this paper. Specifically, our focus is on multiple receivers in a distributed setting, aligning well with the evolving landscape of 6G cell-free systems.



Various studies, such as \cite{Cabric}, have explored sensing location diversity and consensus methods among multiple receivers outside the 6G context. Initiatives like RadioHound \cite{radiohound} and Electrosense \cite{rajendran2017electrosense} highlight the emergence of distributed sensing. However, it's crucial to clarify that while sensing is distributed, the processing remains centralized. 
\Ac{amc} has always been a starting point in the search towards more flexible and cognitive networks and was introduced about 15 years ago as a key technology for wireless networks in several works \cite{haykin,Gardner}. Since then, approaches have moved away from statistical signal processing methods to deep-learning-based models \cite{amc_dl_first}. Extensive research within the domain of \ac{dl}-\ac{amc} has predominantly focused on central models. With the ResNet \cite{oshea} architecture is widely considered one of the most performant feasible architecture. 

Models incorporating multiple inputs and exploring diversity gain have garnered considerable attention, contributing valuable insights to the field \cite{mimo}. However, investigations into fully distributed models with multiple input signals remain relatively scarce. While certain papers, such as \cite{CoAMC}, provide initial insights, a comprehensive understanding of the impact on accuracy and computational complexity is notably limited.  To the best of our knowledge, a research gap exists in conducting in-depth examinations of integrating AMC into distributed cell-free networks, thereby serving as a primary motivation for this paper. This paper addresses this gap by analyzing the impact on performance and total computational load, focusing on two distinct distribution methods for \ac{amc}.

\subsection{Contribution}
This paper proposes, evaluates, and compares two novel approaches for distributed \ac{amc} in cell-free networks, with computationally capable units for the RUs and the DU. The first approach considers a scenario, in which sharing the IQ samples is not feasible due to privacy concerns, fronthaul limitations, or strict synchronization requirements. In this approach, each RU classifies the modulation based on a signal with an SNR that is most likely lower than the combined signal's SNR. The DU predicts the modulation using a voting system of the local soft decision of each RU. Our results show that the distributed model can achieve similar or better results than a central model, where each RU has a model equivalent to the central model. The total computational load scales linearly with the number of considered RUs.  

In the second approach, each RU provides its local soft decision and its IQ samples to the DU. The DU combines the different signals, exploiting diversity gain, and extracts features from the combined signal. The DU makes the prediction based on the RU's local soft decisions and the features extracted from the combined signal. This approach combines the diversity gain from the combined signal with the gain obtained by voting local soft decision. Our results show that this distribution of the computational workload can achieve an increase in classification accuracy for a lower total computational workload.  

The main contribution of this paper is twofold.
\begin{itemize}[leftmargin=*]
\item First, we propose two novel general approaches to implement AMC in cell-free networks, emphasizing the practical implementation and limitations. Exploiting distributed processing on computationally capable units.
\item Second, we propose a practical way to train these distributed approaches to reduce the overall training time and emphasize reusability and scalability. Reducing the training time is based on transfer learning.
\end{itemize}

The remainder of the paper is structured as follows. Section II outlines the network model. Section III introduces the experiment design and the dataset generation. In Section IV, we introduce our classification approaches and training. Subsequently, we present the performance evaluation results in Section V. Finally, the paper concludes with final remarks in Section VI.

\section{Network Model}

\begin{figure}
     \centering
     \begin{subfigure}[b]{0.55\linewidth}
         \centering
         \includegraphics[scale=0.55]{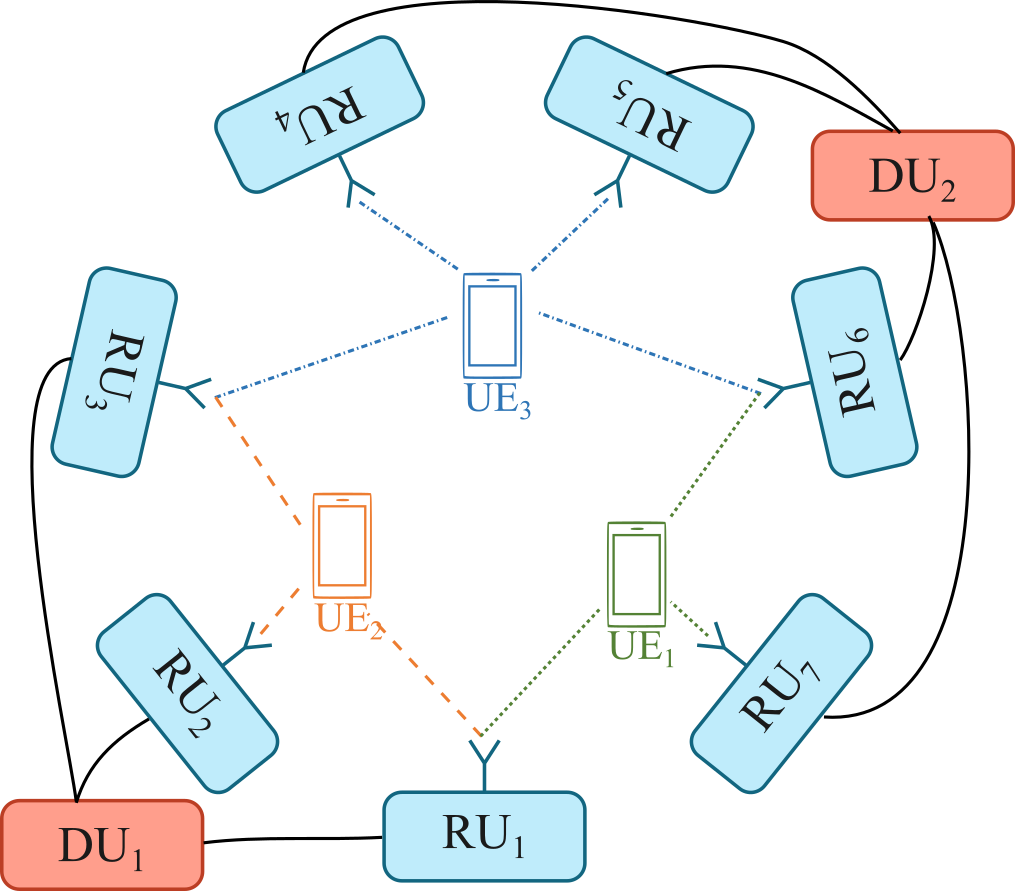}
         \caption{User-centric cell-free network in which multiple RUs serve a UE.}
         \label{fig:systemmodel}
     \end{subfigure}
     \hfill
     \begin{subfigure}[b]{0.4\linewidth}
         \centering
         \includegraphics[scale=0.95]{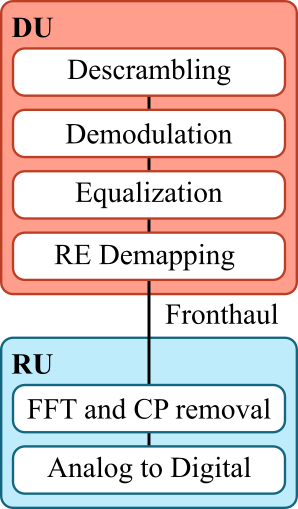}
         \caption{Functionality in uplink with 7.2x functional split.}
         \label{fig:split}
     \end{subfigure}

        \caption{User-centric cell-free network and 7-2x functional split}
        \label{fig:splitted}
        \vspace{-0.3cm}
\end{figure}

The \ac{oran} cell-free network is characterized by a multitude of spatially distributed \acp{ru} engaged in cohesive and collaborative transmission and reception activities, orchestrating the provision of services to \ac{ue} using concurrent time-frequency resources. Multiple \acp{ru} are connected to a \ac{du} through fronthaul links. Initial works (e.g \cite{cellfree2}) assumed a central processing approach in which the \ac{ru} works free of computation, only needing to receive the signals of the \ac{ue} and transfer them to the \ac{du}. However, this approach becomes less practical due to the large computational workload aggregated in the \ac{du}. To this end, distributed processing becomes desirable. By disaggregating the computational workload of the physical layer processing between \acp{ru} and \ac{du}. 

Fig. \ref{fig:systemmodel} shows a user-centric cell-free network with distributed processing. There are multiple ways to split the computational load between the DU and the RU. In the literature, these different splits are referred to as functional splits \cite{functional_splits}. The most commonly considered functional split is the 7-2 split, shown in Fig. \ref{fig:split}. This functional split is followed by the O-RAN specification, making the trade-off between latency requirements and fronthaul bandwidth availability.   
The \ac{ru} primarily handles the low-physical layer processing, encompassing analog-to-digital conversion, digital pre-processing, and amplification of radio signals. On the other hand, the \ac{du} serves as the digital processing counterpart to the \ac{ru}, undertaking functions such as digital beamforming, channel coding, modulation/demodulation, and a spectrum of signal processing operations.  


\section{Experiment design and dataset generation}
\subsection{Experiment design}

This paper considers the uplink in a network of three RUs connected to a single DU and serving the same UE. The UE transmits a signal with a modulation that is not known by the RUs and DU. Before being received by each RU, the signal passes through a channel. This channel applies attenuation and adds white noise. 
After receiving the signal, each RU converts the signal from the analog to the digital domain and applies a \ac{fft} and \ac{cp} removal. The resulting \ac{iq} samples are transmitted to the DU over the fronthaul link. In addition to these functions, each RU may also run a model that predicates the modulation used by the UE. Depending on the use-case, the RU needs information about the modulation locally, or the DU needs to know the modulation centrally. The model running on the RU passes its soft decision to the DU for further processing. At the DU, the \ac{iq} samples of all the different RUs are combined using \ac{egc}. Before performing the demodulation, the DU predicts the modulation using a model running on the DU.

\subsection{Dataset generation}
A synthetic dataset is generated using Matlab to train and evaluate the models considered in this paper. The structure of the dataset generation is based on the structure provided by the experiment design.
Random data is modulated with a modulation type including \ac{bpsk}, \ac{qpsk}, 16 \ac{qam}, 32\ac{qam}, 64\ac{qam}, 128\ac{qam}, and 256\ac{qam}, representing the complex baseband signal transmitted by the UE. This signal is passed through a flat fading channel with attenuation and added white noise. 



The attenuation and white noise added to each RU are calculated such that the \ac{snr} of the \ac{egc} signal is equal to a chosen \ac{snr}. The SNR of the combined signal chosen ranges from $-10dB$ to $30dB$ with a step of $2dB$, resulting in 21 different SNRs.
For each pair of modulation and SNR, we generated 1024 frames of 1024 \ac{iq} samples per RU. To ensure an equal representation of each SNR/Modulation pair, the dataset is split by randomly picking 768 frames for training, 128 frames for validation, and the remaining 128 frames for testing. In total, 150 528 frames were generated for this dataset.



\section{Classification approaches and training}
\begin{figure}
    \centering
    \includegraphics[width=0.7\linewidth]{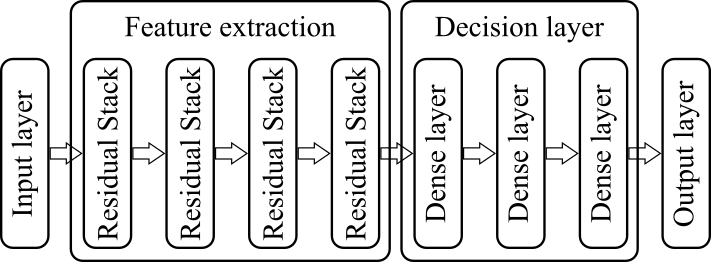}
    \caption{Resnet model with feature extraction of four residual stacks and decision layers.}
    \label{fig:resnet} 
    \vspace{-0.3cm}
\end{figure}


This paper adopts ResNets as the foundation for the models considered in this paper. Fig. \ref{fig:resnet} provides a comprehensive overview of the ResNet model, which is systematically divided into four functional blocks for clarity of discussion. We first introduce the ResNet model and then provide several approaches to map this on a disaggregated cell-free network.

\subsection{General ResNet model}

\begin{figure*}
\begin{subfigure}{0.12\linewidth}
  \centering
  \includegraphics[scale=0.9]{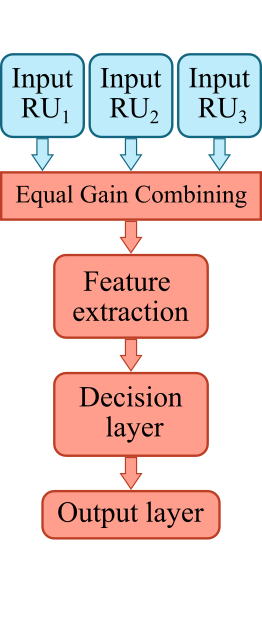}
  \caption{Central model}
  \label{fig:central}
\end{subfigure}
~
\begin{subfigure}{0.32\linewidth}
   \centering
   \includegraphics[scale=0.9]{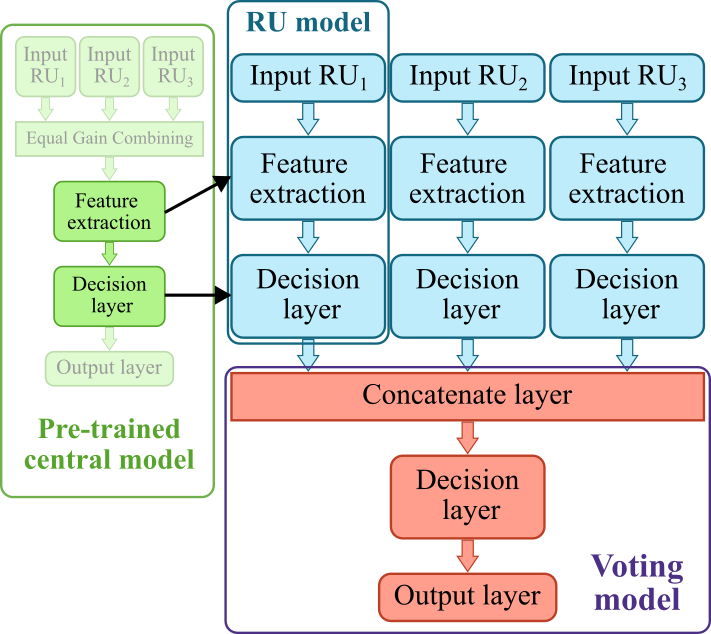}
  \caption{Distributed model}
  \label{fig:distributed}
\end{subfigure}
~
\begin{subfigure}{0.49\linewidth}
   \centering
  \includegraphics[scale=0.9]{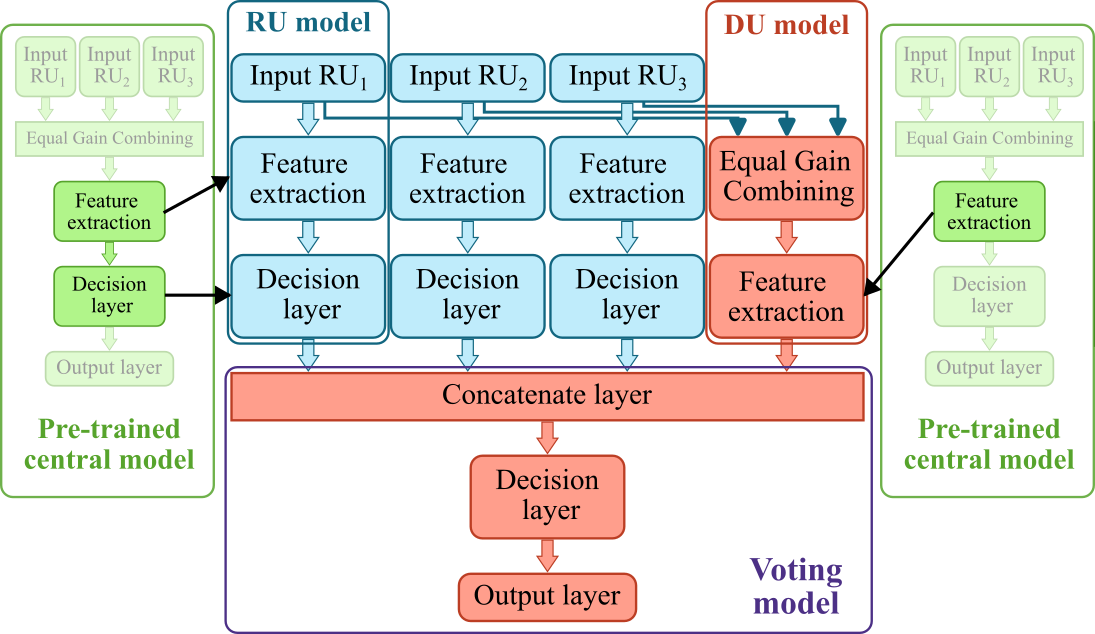}
  \caption{Hybrid model}
  \label{fig:hybrid}
\end{subfigure}

\caption{The baseline and two considered approaches and a schematic of their training process. The blocks in blue represent layers running on the RU, while those in red depict layers running on the DU. The green blocks denote pre-trained models utilized to train the distributed and hybrid models. The black arrows illustrate the transfer of weights between different layers.}
\label{fig:models}
\vspace{-0.4cm}
\end{figure*}

The first block in the ResNet approach, the input layer, interfaces real-world data and the model. The input of this block is the IQ samples derived from the dataset. The output of this block is a vector of dimensions $2 \times N_{IQ}$, with $N_{IQ}$ a parameter controllable when exploiting different variations of the proposed approaches.
The second block, dedicated to feature extraction, transforms the received IQ samples into features via multiple residual stacks. The number of residual stacks is a design parameter, each residual stack consisting of 32 filters with a (3,1) kernel. 
Upon feature extraction, the third block, the decision layer, forms a soft decision. This soft decision exists of the probability with which the decision layer identifies the modulation used in the input IQ samples. The decision layer comprises two dense layers with 128 neurons each and a dense layer with seven neurons (representing each modulation under consideration). The output of the decision layer is considered the soft decision.
Finally, the fourth and last block, the output layer, concludes the model's prediction by selecting the modulation with the highest confidence level. This hierarchical organization of the ResNet model provides a structured framework for understanding its functionality and predictive capabilities.
These four blocks are used in addition to two non-trainable blocks to build three distinct model approaches, shown in Fig. \ref{fig:models}.

\subsection{Baseline Central model} 
The central approach, shown in Fig. \ref{fig:central}, preprocesses the incoming IQ samples from three distinct RUs using \ac{egc}, forming a single input vector of desired dimensions $2 \times N_{IQ}$. The preprocessing
happens in a non-trainable EGC block, introducing diversity gain by enhancing the \ac{snr}. This EGC block sums the IQ samples of each distinct RU, requiring strict synchronization in order to achieve the most diversity gain. 

The resultant combined signal serves as the input for the subsequent Feature Extraction block, closely mirroring the structure of the original ResNet model. 
The training spans 10 epochs using the train part of the dataset, implementing early stopping based on validation data. The model exhibiting the highest validation accuracy is selected for testing.
This model is deemed central, as all computations are executed at the DU. This central model will be used as the baseline against which the distributed and hybrid models are compared.






\subsection{Distributed model}
The distributed model, illustrated in Fig. \ref{fig:distributed}, distributes the computational workload among individual RUs, minimizing the computational burden on the DU. It consists of two distinctive sub-models: the RU-model and the voting-model. The RU-model functions independently on each RU, producing a soft decision based on the incoming IQ samples specific to that RU. These soft decisions are relayed to the DU. The DU, employing the voting-model, predicts the modulation by concatenating the soft decision from each RU using a non-trainable concatenate layer, followed by a decision and an output layer. No IQ sample transmission is needed between the DU and the RUs, resulting in both lower fronthaul capacity and synchronization requirements.

The training process unfolds in two phases, each corresponding to one of the sub-models. The first phase revolves around the RU-model, which undergoes training through transfer learning. For this purpose, a central model with an equivalent input size and number of residual layers is trained as targeted in the RU-model. Once this training is complete, the weights of the feature extraction layer and the decision layer from the central model are transferred to the corresponding layers of the RU-model. Following this transfer, all weights of the RU-model are fixed, rendering them non-trainable for the subsequent phase. This fixed RU-model is then duplicated and deployed across every RU. 

In the second phase, the voting-model undergoes training, utilizing the soft decision provided by the RUs. This training strategy is flexible and reusable for setups with different numbers of RUs. The RU-model is fixed and can be duplicated; the voting-model must be retrained with changing amounts of RUs. 
Each phase of the training spans 10 epochs using the same train part, validation part, and test part of the dataset, implementing early stopping based on validation data. The model exhibiting the highest validation accuracy is selected for testing.

\subsection{Hybrid model}
In the hybrid model, each RU provides its IQ samples and soft decisions to the DU. The hybrid model extends the distributed model by introducing an additional sub-model, namely, the DU-model. The hybrid model with the three sub-models is depicted in Fig. \ref{fig:hybrid}. The DU-model combines the IQ samples from different RUs with a non-trainable \ac{egc} block, and features are extracted using the resultant combined IQ samples, which are then forwarded to the concatenate layer of the voting-model. Unlike the distributed model, where only the soft decisions from individual RU-models contribute to the prediction, the voting-model in the hybrid model also integrates features extracted by the DU-model. 
The RU-model and DU-model can have varying input sizes and numbers of residual stacks. In this model, a difference in input size can be achieved by clipping the signal before feature extraction.

Training the hybrid model involves three phases. In the first phase, the RU-model undergoes training similar to the RU-model of the distributed model. Initially, a central model is trained, and subsequently, the weights of the corresponding layers are transferred between the pre-trained central model and the RU-model. Post-training, these weights are fixed, and the RU-model is duplicated to operate on each RU.  
The second phase, occurring concurrently, focuses on the DU-model. For this, a central model is trained with the same input size and the number of residual stacks as desired for the feature extraction layer of the DU-model. After training, the weights are transferred and fixed.
Finally, the voting-model is trained using soft decisions from the different RUs and features extracted by the DU-model. In the hybrid model, both the RU-model and the DU-model operate independently, providing high flexibility and reusability. When the number of RUs increases, only the voting-model needs retraining.

\section{Results}
To assess and compare the different proposed distributed approaches, we define data scenarios where the mean SNR across all RUs equals the EGC-SNR divided by the number of RUs. In these scenarios, the performance gains achieved are not solely determined by overall SNR increases but rather stem from the organization of data processing and effective utilization of SNR diversity. 
Fixing the mean SNR across all RUs allows us to conveniently assess the impact of scalability, which involves introducing more RU-level decisions into the problem space. Each additional RU contributes incremental SNR, albeit at the expense of increased computational demands. Scalable system implementations ensure that implementation costs scale gracefully with network size. Performance scaling depends on how information is utilized and how diversity is leveraged.

To quantitatively evaluate the performance of the proposed distributed approaches, we define two key metrics: model accuracy and computational efficiency.
\begin{itemize}[leftmargin=*]
\item \textit{Model accuracy} is measured as the percentage of correctly classified modulation schemes. Higher accuracy indicates better performance. To account for randomness, accuracy is averaged over 16 Monte Carlo simulations.

\item \textit{Computational Efficiency} is assessed by the number of \acp{flop} required for modulation classification. A lower FLOP count signifies greater computational efficiency. TensorFlow's built-in estimator is employed to estimate the FLOPs for each model. However, it is essential to note that these estimates are approximations and may not reflect the exact computational requirements.
\end{itemize}
In the results, we first discuss the central model in detail. The central model is the foundation for both the distributed and hybrid models, making it a crucial component of our analysis. Insights from examining the central model's performance are valuable in streamlining the search space for the two proposed approaches.

\subsection{Baseline model: Central model}
The entire search space is considered to analyze the central model, which combines the signals of three RUs. The feature extraction layer's residual stacks range from 4 to 7, and the model's input size varies between 128, 256, 512, and 1024 IQ samples. \acp{mflop} and average accuracy are calculated for each pairing of input size and residual stacks.

\begin{table}[]
\caption{The optimal parameters for the baseline and distributed model }
    \label{tab:cent_distributed_overview}
    \centering
\begin{tabular}{rrrrrr}
    \multicolumn{2}{l}{}                                    &          \multicolumn{2}{c}{Centralized}                    &          \multicolumn{2}{c}{Distributed}                 \\ 
    \multicolumn{2}{l}{}                                    &          \multicolumn{2}{c}{\scriptsize with strict synchronization }                    &          \multicolumn{2}{c}{\scriptsize without synchronisation }                 \\ \cline{3-6} 
    \multicolumn{1}{c}{\#IQ}  &\multicolumn{1}{c}{\#Stacks} &\multicolumn{1}{|c}{\#MFLOPS}&\multicolumn{1}{c||}{Accuracy} &\multicolumn{1}{c}{\#MFLOPS}&\multicolumn{1}{c|}{Accuracy}\\ \hline \hline
    \multicolumn{1}{|c|}{128} &\multicolumn{1}{c}{5}        &\multicolumn{1}{||c|}{3.23}  &\multicolumn{1}{c||}{55.92\%}  &\multicolumn{1}{c|}{9.72}   &\multicolumn{1}{c|}{55.91\%} \\ \hline
    \multicolumn{1}{|c|}{256} &\multicolumn{1}{c}{7}        &\multicolumn{1}{||c|}{6.56} &\multicolumn{1}{c||}{58.65\%}  &\multicolumn{1}{c|}{19.79}  &\multicolumn{1}{c|}{58.17\%} \\ \hline
    \multicolumn{1}{|c|}{512} &\multicolumn{1}{c}{4}        &\multicolumn{1}{||c|}{12.53} &\multicolumn{1}{c||}{60.74\%}  &\multicolumn{1}{c|}{37.60}  &\multicolumn{1}{c|}{60.15\%} \\ \hline
    \multicolumn{1}{|c|}{1024}&\multicolumn{1}{c}{5}        &\multicolumn{1}{||c|}{25.77} &\multicolumn{1}{c||}{61.92\%}  &\multicolumn{1}{c|}{77.32}  &\multicolumn{1}{c|}{61.51\%} \\ \hline
\end{tabular}
\vspace{-0.3cm}
\end{table}

Table \ref{tab:cent_distributed_overview} shows the best pairing for each input size in combination with the number of residual layers resulting in the highest accuracy. As shown in other research studies, the model's accuracy depends on the input size; a higher input size results in higher accuracy. There is no distinct pattern regarding the optimal number of layers for each input size. Comparing the computational complexity for all the different pairings of parameters, we find that increasing the number of residual stacks marginally impacts the FLOP count; input size plays a more prominent role in determining computational complexity. This phenomenon is attributed to the predominant number of FLOPs in calculating the initial and second residual stacks. Conversely, the input size substantially influences FLOPs, establishing it as the primary determinant for reducing computational complexity instead of the number of layers. The central model achieves the highest accuracy with an input size of 1024 and five residual stacks, although this also requires the highest computational complexity.

\subsection{Distributed model}

As the distributed model is built upon a single central model and transfer learning for the RU-model, the search space is the same as the search space of the central model. We consider a distributed model with three identical RU-models, allocating the computational load equally across the RUs, demanding only minimal computational resources from the DU.

Examining optimal combinations for the central case is adequate to identify the most accurate model for the distributed scenario. Table \ref{tab:cent_distributed_overview} shows accuracy and its associated computational complexity. The MFLOPs for the distributed model can be estimated by aggregating the MFLOPs of individual RU models. Similar to the central network, a pronounced correlation exists between input size and MFLOPs in the distributed network.

The accuracy of distributed models with three RU-models aligns closely with the accuracy achieved by a central model with an equivalent input size and number of residual stacks. This shows that the diversity gained from the distributed model aligns with the EGC's, rendering it an appealing choice when EGC is not feasible. Regrettably, the cost of this diversity gain is threefold compared to that of the central model.

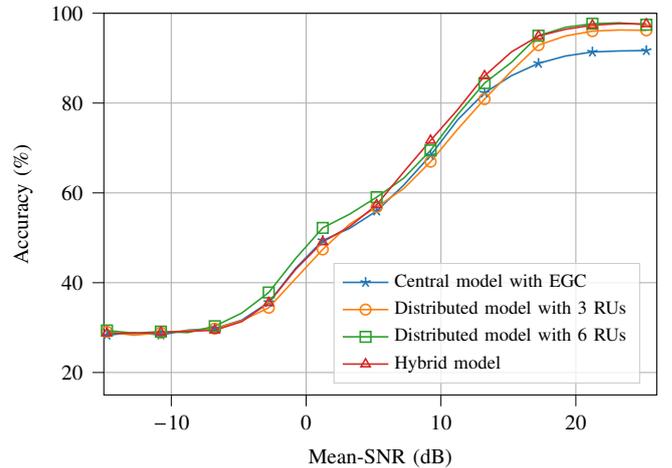
\begin{figure}
    \centering
\begin{tikzpicture}

\definecolor{crimson2143940}{RGB}{214,39,40}
\definecolor{darkgray176}{RGB}{176,176,176}
\definecolor{darkorange25512714}{RGB}{255,127,14}
\definecolor{forestgreen4416044}{RGB}{44,160,44}
\definecolor{lightgray204}{RGB}{204,204,204}
\definecolor{steelblue31119180}{RGB}{31,119,180}

\begin{axis}[
legend cell align={left},
legend style={
  fill opacity=0.8,
  draw opacity=1,
  text opacity=1,
  at={(0.97,0.03)},
  anchor=south east,
  draw=lightgray204,
  font=\scriptsize,
},
tick align=outside,
tick pos=left,
x grid style={darkgray176},
xlabel={Mean-SNR (dB)},
xmajorgrids,
xmin=-15, xmax=26,
xtick style={color=black},
y grid style={darkgray176},
ylabel={Accuracy (\%)},
ymajorgrids,
ymin=15, ymax=100,
ytick style={color=black},
width=\linewidth,
height=0.75\linewidth,
]
\addplot [semithick,mark=star,mark repeat=2,mark phase = 1, steelblue31119180]
table {%
-14.7712125471966 28.3203125
-12.7712125471966 28.8853236607143
-10.7712125471966 28.369140625
-8.77121254719662 29.443359375
-6.77121254719662 29.8130580357143
-4.77121254719662 31.6755022321429
-2.77121254719662 35.6096540178571
-0.771212547196624 43.2477678571429
1.22878745280338 49.3931361607143
3.22878745280338 52.1065848214286
5.22878745280338 55.9640066964286
7.22878745280338 61.71875
9.22878745280338 68.359375
11.2287874528034 76.2555803571428
13.2287874528034 82.275390625
15.2287874528034 86.1049107142857
17.2287874528034 88.8462611607143
19.2287874528034 90.478515625
21.2287874528034 91.3783482142857
23.2287874528034 91.6085379464286
25.2287874528034 91.69921875
};
\addlegendentry{Central model with EGC}
\addplot [semithick, mark=o,mark repeat=2,mark phase = 1,darkorange25512714]
table {%
-14.7712125471966 29.052734375
-12.7712125471966 28.2993861607143
-10.7712125471966 28.8783482142857
-8.77121254719662 29.3178013392857
-6.77121254719662 29.7572544642857
-4.77121254719662 31.5359933035714
-2.77121254719662 34.423828125
-0.771212547196624 40.9388950892857
1.22878745280338 47.412109375
3.22878745280338 53.0412946428572
5.22878745280338 56.8359375
7.22878745280338 60.9305245535714
9.22878745280338 66.943359375
11.2287874528034 74.1559709821428
13.2287874528034 80.8872767857143
15.2287874528034 87.1303013392857
17.2287874528034 92.9059709821429
19.2287874528034 94.9428013392857
21.2287874528034 96.0170200892857
23.2287874528034 96.2751116071428
25.2287874528034 96.1983816964286
};
\addlegendentry{Distributed model with 3 RUs }
\addplot [semithick, forestgreen4416044,mark=square,mark repeat=2, mark phase=1]
table {%
-14.7712125471966 29.3666294642857
-12.7712125471966 28.7667410714286
-10.7712125471966 29.1155133928571
-8.77121254719662 28.8922991071429
-6.77121254719662 30.3152901785714
-4.77121254719662 33.2449776785714
-2.77121254719662 37.8208705357143
-0.771212547196624 45.4241071428571
1.22878745280338 52.2042410714286
3.22878745280338 55.2734375
5.22878745280338 59.0401785714286
7.22878745280338 63.2533482142857
9.22878745280338 69.4614955357143
11.2287874528034 77.4135044642857
13.2287874528034 84.4447544642857
15.2287874528034 89.0764508928571
17.2287874528034 95.0055803571429
19.2287874528034 96.875
21.2287874528034 97.6702008928571
23.2287874528034 97.8794642857143
25.2287874528034 97.4469866071429
};
\addlegendentry{Distributed model with 6 RUs}
\addplot [semithick,  mark=triangle,mark repeat=2,mark phase = 1,crimson2143940]
table {%
-14.7712125471966 28.7737165178571
-12.7712125471966 28.7039620535714
-10.7712125471966 28.9132254464286
-8.77121254719662 29.1713169642857
-6.77121254719662 29.4084821428571
-4.77121254719662 31.3406808035714
-2.77121254719662 35.5259486607143
-0.771212547196624 42.9896763392857
1.22878745280338 49.0792410714286
3.22878745280338 52.5111607142857
5.22878745280338 57.3451450892857
7.22878745280338 64.6414620535714
9.22878745280338 71.6866629464286
11.2287874528034 78.5086495535714
13.2287874528034 86.0491071428572
15.2287874528034 91.40625
17.2287874528034 94.9288504464286
19.2287874528034 96.4285714285714
21.2287874528034 97.3353794642857
23.2287874528034 97.6841517857143
25.2287874528034 97.6213727678571
};
\addlegendentry{Hybrid model}
\end{axis}

\end{tikzpicture}
    \caption{Accuracy vs Mean-SNR for a central model with EGC, a distributed model with 3 RUs, and a distributed model with 6 RUs with input size 512, and a hybrid model with input size 256 for the DU and 128 for the RU model}
    \label{fig:dist_accsnr}
    \vspace{-0.3cm}
\end{figure}

While the central model doesn't exhibit performance improvement with adding more RUs, the distributed model demonstrates substantial gains in accuracy as shown in Fig. \ref{fig:dist_accsnr}. This figure depicts the accuracy as a function of mean SNR for the central model with EGC, the distributed model with 3 RUs, and the distributed model with 6 RUs. All models have an input size of 512 and 7 residual stacks. The central and distributed models with 3 RUs perform similarly for SNRs up to $12dB$. Beyond this SNR, the central model's performance deteriorates. The distributed model with 6 RUs consistently outperforms both other models. However, this superior performance comes at the cost of six times the computational cost of the central model.

The distributed model offers a viable alternative approach when EGC is not feasible. It can achieve comparable accuracy to the central model while maintaining decentralized processing. However, this distributed approach necessitates a trade-off between computational efficiency and accuracy. While it eliminates strict synchronization requirements posed by EGC, it incurs a substantial computational cost, which scales linearly with the number of RUs. 

\subsection{Hybrid model}
In the hybrid model, two distinct central models are considered when training the RU-model and DU-model. To this end, there are four design parameters to consider. Based on the analysis performed on the central model, we can eliminate two design parameters as an optimal number of residual stacks is discerned for a given input size. The best-performing pairs of input size and number of residual stacks identified during the central model analysis are used. We consider a hybrid model with identical RU-models and a distinct DU-model, where the computational load is distributed across the RUs and the DU.


Since the primary computational load is in the feature extraction of both the RU-model and DU-model, the computational load at the DU and RU can be approximated by aggregating the MFLOPs of the central model trained for the DU-model and RU-model, respectively.
Table \ref{tab:acc_hybrid} shows the accuracy of different combinations of RU input sizes and DU input sizes. The accuracy experiences significant improvement with a larger input size for the DU-model, while changing the RU-model input size does not show any improvements.
\begin{table}[]
\caption{Accuracy of the hybrid model with varying input size for the DU and RU model}
    \label{tab:acc_hybrid}
    \centering
\begin{tabular}{rllll}
    \multicolumn{1}{l}{}       & \multicolumn{4}{c}{\# IQ DU}                                                            \\ \cline{2-5} 
    \multicolumn{1}{c|}{\#IQ RU}  & \multicolumn{1}{c|}{128} & \multicolumn{1}{c|}{256} & \multicolumn{1}{c|}{512} & \multicolumn{1}{c|}{1024} \\ \hline
    \multicolumn{1}{|r|}{128}  & \multicolumn{1}{l|}{59.30\%}  & \multicolumn{1}{l|}{61.74\%}  & \multicolumn{1}{l|}{62.87\%}  & \multicolumn{1}{l|}{64.16\%}  \\ \hline
    \multicolumn{1}{|r|}{256}  & \multicolumn{1}{l|}{60.99\%}  & \multicolumn{1}{l|}{61.66\%}  & \multicolumn{1}{l|}{62.91\%}  & \multicolumn{1}{l|}{64.24\%}  \\ \hline
    \multicolumn{1}{|r|}{512}  & \multicolumn{1}{l|}{61.77\%}  & \multicolumn{1}{l|}{62.96\%}  & \multicolumn{1}{l|}{62.84\%}  & \multicolumn{1}{l|}{64.41\%}  \\ \hline
    \multicolumn{1}{|r|}{1024} & \multicolumn{1}{l|}{62.7\%}  & \multicolumn{1}{l|}{ 63.7\%}  & \multicolumn{1}{l|}{63.36\%}  & \multicolumn{1}{l|}{64.30\%}  \\ \hline
\end{tabular}
\vspace{-0.3cm}
\end{table}

Increasing the input size for the DU-model significantly enhanced the accuracy of the hybrid model, whereas increasing the input size for the RU-model failed to produce comparable improvements.  Increasing the RU-model's input size substantially elevates the overall computational burden of the hybrid model due to the integration of multiple RU-models. A direct comparison between a centralized model with an input size of 1024 and a hybrid model with input sizes of 1024 for the DU-model and 128 for the RU-model unveiled a notable $2.5\%$ gain for the hybrid model.

Fig. \ref{fig:dist_accsnr} illustrates the comparison between a central model with an input size of 512 and a hybrid model with a DU input size of 256 and an RU input size of 128. Both models require a similar total computational load of 12.53 MFLOPs for the central model and 16.30 MFLOPs for the hybrid model. The hybrid model consistently outperforms the central model in accuracy across various SNRs. The hybrid model combines the diversity gain from EGC and the voting gain. When EGC is available at the DU, the hybrid model should be considered as it outperforms in terms of accuracy with an equivalent but distributed total computational load. This effect is more apparent when comparing a centralized model with an input size of 1024 and a hybrid model with input sizes of 512 for the DU-model and 128 for the RU-model. In this case, the hybrid model outperforms the central model both in terms of computational complexity (22.24 MFLOPS vs 25.77 MFLOPS) and accuracy ($62.87\%$ vs $61.92\%$).

\section{Conclusion and future work}
In this paper, we introduced two novel distributed approaches for AMC in cell-free networks. We evaluate these models based on their accuracy and computational complexity. The distributed model addresses scenarios where IQ sample sharing is impractical due to fronthaul constraints or privacy considerations. Compared to a central model with EGC, which is an upper bound that requires tight synchronization between RUs, the distributed model achieves similar complexities without the synchronization constraint. However, the computational demands of the distributed model are several times higher than those of the central model. 
When EGC is possible at the DU, the hybrid model strikes a balance between accuracy and complexity, offering efficient workload distribution and surpassing the central model.

Our work establishes a foundational framework for AMC in cell-free networks, opening up new research avenues. These include exploring more intricate channels beyond the AWGN channel, investigating alternative signal-combining methods for RUs beyond EGC, and exploring distributed AMC with architectures beyond ResNet.



\bibliographystyle{ieeetr}
\bibliography{bibliography}

\end{document}